\documentclass[preprint,preprintnumbers,amsmath,amssymb]{revtex4}
\usepackage{epsfig}
\usepackage{amssymb}
\usepackage{bm}

\usepackage{latexsym}
\usepackage{amsmath}
\usepackage{slashed}
\usepackage{graphics}
\usepackage{epsfig}
\def\br{\begin{eqnarray}}
\def\er{\end{eqnarray}}
\def\be{\begin{equation}}
\def\ee{\end{equation}}

\def\({\left(}
\def\){\right)}
\def\<{\left\langle}
\def\>{\right\rangle}

\begin{document}
\title{The Extreme Walking Behavior in a 331-TC Model}
\author{A. Doff}
\email{agomes@utfpr.edu.br}
\affiliation{Universidade Tecnol\'ogica Federal do Paran\'a - UTFPR - DAFIS
Av Monteiro Lobato Km 04, 84016-210, Ponta Grossa, PR, Brazil }

\date{\today}

\begin{abstract}
It is quite possible that the Technicolor problems are related to the poorly known self-energy expression, or the way chiral symmetry breaking (CSB) is realized in non-abelian gauge theories. Actually, the only known laboratory to test the CSB mechanism is QCD. The TC dynamics may be quite different from the QCD , this fact has led to the walking TC proposal making the new strong interaction almost conformal and changing appreciably its dynamical behavior.  There are  different ways to obtain  of extreme walking (or quasi-conformal) technicolor theories, in this paper we propose an scheme to obtain this behavior based on an extension of the electroweak sector of the standard model,  in the context of so called 331-TC model. 
\end{abstract}

\pacs{12.60.Cn, 12.60.Rc, 11.30.Na}

\maketitle

\section{Introduction}

\par The $125$ GeV new resonance discovered at the LHC \cite{LHC} has many of the characteristics expected for the Standard Model (SM) Higgs boson. If this particle is a composite or an elementary scalar boson is still an open question that probably will be answered in the next LHC run. In recent papers the ATLAS  and CMS Collaborations\cite{diboson} reported an experimental anomaly in diboson
production with apparent excesses in $WW$, $WZ$  and $ZZ$ channels and  this anomaly have   inspired a number of theoretical papers  proposing as an explanation  the production of heavy weak bosons,  $W'$ and $Z'$ . 
\par Thus, it becomes interesting to investigate the possibility of obtaining a   light scalar boson in the context of  models which features contributions from new heavy weak bosons $W'$ and $Z'$.  In some  extensions of the standard model (SM), as in the so called 3-3-1 models\cite{3m1} $SU(3)_{{}_{L}}\otimes SU(3)_{{}_{c}} \otimes U(1)_{{}_{X}}$,  new massive neutral and charged gauge bosons, $Z'$  and  $V^{\pm} $, are predicted.   The 3-3-1 model is the minimal gauge group that at the leptonic level admits charged fermions and their antiparticles as members of the same multiplet,  the predictions of the $GW = SU(2)_{{}_{L}}\otimes U(1)_{{}_{Y}}$ alternative models are leptoquark fermions with electric charges $±5/3$ and $4/3$ and bilepton gauge bosons with lepton number $L = \pm 2$. The quantization of electric charge is inevitable in the $G=3m1$ models\cite{QQ} with three non-repetitive fermion generations brea-king generation universality and does not depend on the character of the neutral fermions.
\par  In the Ref.\cite{Das}  was suggested that the gauge symmetry breaking of a specific version of a 3-3-1 model\cite{3m1} would be implemented dynamically because at the scale of a few TeVs the $U(1)_X$ coupling constant becomes strong and the  exotic quark $T$(charge $5/3$)  will form a  $U(1)_X$ condensate breaking  $SU(3)_{{}_{L}}\otimes U(1)_X$ to the electroweak symmetry.  This possibility was explored by us in the Ref.\cite{331us} assuming a model based on the gauge symmetry  $SU(2)_{TC}\otimes SU(3)_{{}_{L}}\otimes SU(3)_{{}_{c}} \otimes U(1)_{{}_{X}}$  , where the electroweak symmetry is broken dynamically by a technifermion condensate, that is characterized by the $SU(2)_{TC}$ Technicolor (TC) gauge group.  The early technicolor models \cite{weinberg}  suffered from problems like flavor changing neutral currents (FCNC) and contributions to the electroweak corrections not compatible with the experimental data, as can be seen in the reviews of Ref.\cite{tc}. However, the TC dynamics may be quite different from the known strong interaction theory, i.e. QCD, this fact has led to the walking TC proposal \cite{walk}, which are theories where the incompatibility with the experimental data has been sol-ved, making the new strong interaction almost conformal and changing appreciably its dynamical behavior.
\par  We can obtain an almost conformal TC theory, when the fermions are in the fundamental representation, introducing a large number of TC fermions ($n_{TF}$), leading to an almost zero $\beta$ function and flat asymptotic coupling constant. The cost of such procedure may be a large S parameter\cite{peskin}  incompatible with the high precision electroweak measurements. However, this problem can be solved by assuming that TC fermions are in other representations than the funda-mental\cite{sannino1} and an effective Lagrangian analysis indicates that such models also imply in a light scalar Higgs boson \cite{sannino2}. This possibility was also investigated and confirmed  through the use of an effective potential for composite operators \cite{us1} and through a calculation involving the Bethe-Salpeter equation (BSE) for the scalar state \cite{us2}.    
\par The reason for the existence of the different models (or different potentials) for a composite scalar boson, is a consequence of our poor knowledge of the strongly interacting theories, that is reflected in the many choices of parameters in the effective potentials.  The possibility of obtaining a light composite scalar  according to the approach discussed in Ref.\cite{us1}, is that this result is a direct consequence  of extreme walking (or quasi-conformal) technicolor theories, where the  asymptotic self-energy  behavior  is described by Irregular form of TC fermions\cite{us1,us2}\footnote{In the Eq.(1) $\mu$ is the characteristic scale of mass generation of the theory forming the composite  scalar boson,  $b$ is the coefficient of the $g^3$ term in the renormalization group $\beta(g)$ function, $\gamma = \frac{3c}{16\pi^2 b}$ and  c is the quadratic Casimir operator given by $ c = \frac{1}{2}(C_2(R_1) + C_2(R_2)-C_2(R_3) )$ where $C2(R_i)$ are the Casimir operators for fermions in the representations $R_1$ and $R_2$ that form a composite boson in the representation $R_3$.} 
\be
\Sigma^{(0)} (p^2) \sim \mu \left[1 + bg^2 (\mu^2) \ln\left(p^2/\mu^2 \right) \right]^{-\gamma} \,. 
\ee 
\par In the Ref.\cite{twoscale} we considered the possibility of a light composite scalar boson  arising from mass mixing between a relatively light and  heavy scalar singlets from a see-saw mechanism  expected to occur in two-scale Technicolor (TC) models and we identified that, regardless of the approach used for generating a light composite scalar boson, the behavior exhibited by extreme walking technicolor theories,  is the main feature needed to produce a light composite scalar boson compatible with the boson observed at the LHC. 
\par After this brief motivation of the importance of extreme walking behavior to generate a light composite scalar boson in TC models, in addition to  possibility of 331-TC model contain the necessary requirements to explain the anomaly in diboson production,  in this paper we propose an scheme to obtain the quasi-conformal behavior based on an extension of the electroweak sector of the standard model,  331-TC model ($SU(N)_{TC}\otimes [SU(3)_{{}_{L}}\otimes SU(3)_{{}_{c}}\otimes U(1)_{{}_{X}}]$). In this model only exotic techniquarks ($U', D'$) will acquire  dynamically generated  mass due $U(1)_X$ interaction at $\Lambda_{{}_{331}}\!\!\sim\! O(TeV)$ .  The terminology {\it exotic} refers to nomenclature used in $331$ models  to designate the allocation of fractional charges assigned to the new quarks ($T, D$). In analogy with this nomenclature ($U', D'$)  are termed  "exotic techniquarks", and all techniquarks  are in the  fundamental representationn, $(R = F)$,  of $SU(N)_{TC}$.
\par  Technicolor models  with fermions in the fundamental representation  are subject to strong experimental constraints that comes from the limits on the $S$ parameter. In our case,  the contribution due to the  TC sector should still lead to a value to the S parameter compatible with the experimental data.  At low energies, i.e.  at the scale associated with electroweak symmetry breaking, we should only consider the contribution of four techniquarks because (U 'and D') are singlets of $SU(2)_{L}$ and do not contribute directly to the bosons (W and Z) masses
\par The exotic technifermions  will present the extreme walking behavior, the usual techniquarks ($U, D$) will present the known asymptotic self-energy behavior  predicted by the operator product expansion(OPE)\cite{ope}. The exotic techniquarks   will have two different  energy scales and the 331-TC model corresponds to an example of  two-scale Technicolor (TC) model. This  article is organized as follows: In section II we present the $U(1)_{X}$ contribution to fermions and technifermions self-energy, in section III we compute the dynamically generated masses to heavy  exotic quarks $(T,S,D)$ and techniquarks $(U', D')$ , where we reproduce the results obtained in Ref.\cite{Das} for heavy  exotic quarks. In the  section IV we illustrate how to obtain the extreme walking behavior in the context of  331-TC model and  Section V contains our conclusions. 

\section{ The $U(1)_{X}$ Contribution to Fermions and Technifermions Self-Energy}

 \par As described in Refs.\cite{Das,331us} the gauge symmetry breaking in 3-3-1 models can be implemented dynamically because at the scale of a few TeVs  the $U(1)_X$ coupling constant becomes strong.  The exotic quark $T$  introduced  will form a condensate  breaking  $SU(3)_{{}_{L}}\otimes U(1)_X$ to  electroweak symmetry, in this paper we will numerically determine the $U(1)_X$ contribution to dynamic mass of  exotic quarks $(T,S,D)$ and  techniquarks $(U', D')$ that appear in the   fermionic content  of the model\cite{331us}  following the same procedure described in Ref.\cite{Das}. The Schwinger-Dyson equation  for quarks(techni-quarks)  due to $U(1)_{X}$ interaction can be written as\cite{Das,331us}
\begin{equation}
S^{-1}(p) = \not{\!\!p} -i\int\frac{d^4q}{(2\pi)^4}\Gamma_{\mu}(p,q)S(q)\Gamma_{\nu}D^{\mu\nu}_{{}_{M_{Z'}}}(p-q)
\label{sde}
\end{equation}
\noindent where in the equation  above we assumed the rainbow approximation for the vertex $\Gamma_{\mu, \nu}$, with 
$\Gamma_{\mu,\nu} = (g_{{}_{V}}\gamma_{\mu,\nu} - g_{{}_{A}}\gamma_{\mu,\nu}\gamma_{5})$, $g_{{}_{V}} = g^2_{X}(Y_{{}_{L}} + Y_{{}_{R}})/4$ and $g_{{}_{A}} = g^2_{X}(Y_{{}_{L}} - Y_{{}_{R}})/4$, where $Y_i$ are  $U(1)_{{}_{X}}$ hypercharges attributed at chiral components of the exotic quarks(techniquarks). 
\par With the purpose of simplifying the calculations it is convenient to choose the Landau Gauge. In this case the $Z'$ propagator  can be written in the following form 
$$
iD^{\mu\nu}_{{}_{M_{Z'}}}(p - q) = -i\frac{\left[g_{\mu\nu} - (p-q)_{\mu}(p - q)_{\nu}/(p-q)^2\right]}{(p - q)^2 - M^2_{Z'}}.
$$
\noindent Writing the  quark propagator  as $i{S}_{{}_{F}}^{\,-1}(p) = i(\slashed{p} - \Sigma_{X}(p^2))$, and considering the equation above, we finally can write  in the  euclidean space the gap equation for $\Sigma(p^2)$\cite{Das}  
\begin{eqnarray}
\Sigma_{X}(p^2) =  && a\int_{0}^{p^2} d^2q q^2 \frac{\Sigma_{X}(q^2)}{[q^2 + \Sigma^2_{X}(q^2)]}\frac{1}{[p^2 + M^2_{Z'}]} +\nonumber \\
                   && a\int_{p^2}^{\Lambda^2} d^2q q^2 \frac{\Sigma_{X}(q^2)}{[q^2 + \Sigma^2_{X}(q^2)]}\frac{1}{[q^2 + M^2_{Z'}]}
\end{eqnarray}
\noindent where $a = \frac{g^2_{{}_{X}}Y_{{}_{L}}Y_{{}_{R}}}{16\pi^2} = \beta Y_{{}_{L}}Y_{{}_{R}}$. To obtain the last equation  we  assumed the angle approximation to transform the term \\ $\frac{1}{(p - q)^2 + M^2_{Z'}}$ as
\br 
\frac{1}{(p - q)^2 + M^2_{Z'}} = \frac{\theta(p - q)}{p^2 + M^2_{Z'}} + \frac{\theta(q - p)}{q^2 + M^2_{Z'}}. 
\er 
 \noindent The integral equation  described above can be transformed into a differential equation for $f(x)$  introducing  the  new variables $x = \frac{p^2}{M^2}$,   with $f(x) = \frac{\Sigma_{X}(x)}{M}$ and  $\alpha=\frac{M_{Z'}}{M}$ that we  reproduce below 
\be 
f''(x) + \frac{2}{x + \alpha^2}f'(x) + \frac{\beta Y_{{}_{L}}Y_{{}_{R}}}{(x + \alpha^2)^2}\frac{xf(x)}{(x + f^2(x))} = 0, 
\ee 
\noindent where $M \equiv\Sigma_{X}(0)$ is the dynamical mass of exotic quarks(or techniquarks) generated  by $U(1)_X$ interaction and the respective boundary conditions for $f(x)$ are  $f(0)=1$ and $f'(0)\\=0$.  In order to obtain the mass spectrum generated due to  $U(1)_X$ interaction, we follow the same procedure described in Ref.\cite{Das} where the $(Y_i)$ hipercharges of exotic quarks $(T,S,\\ D)$ were assumed according to the table 1 and we include the corresponding hipercharges of exotic techniquarks $(U',D')$
\cite{331us}  that are singlets of $SU(3)_c$.

\begin{table}
\centering
\caption{$(Y_i)$ Hipercharges of exotic quarks $(T,S,D)$ and  exotic techniquarks $(U',D')$.}
\label{parset}
\begin{tabular*}{\columnwidth}{@{\extracolsep{\fill}}llll@{}}
\hline
$Y_L$ & $Y_R$ & Exotic Fermion  & Charge\\
      &       & (Technifermion) &        \\
\hline
$-4/3$  &  $-10/3$       & T   & $+5/3$ \\ 
$+2/3$  &  $+8/3$        & D,S & $-4/3$ \\ 
$-1$  &    $-3$         & U'  &  $+3/2$\\ 
$+1$   &   $+3$         & D'  &  $-3/2$\\
\hline
\end{tabular*}
\end{table} 

\section{Dynamically generated masses of exotic quarks and techniquarks due $U(1)_{X}$ contribution}

\par As commented in the previous section we follow the same procedure described in Ref.\cite{Das}, therefore, in order to get an estimate of the $U(1)_X$ dynamically generated  mass, for exotic fermions(or technifermions),   we will numerically solve the Eq.(3) imposing an ultraviolet
cutoff $\Lambda$  on this equation. If the gap equation accepts a $M_T$ solution($M_T$  is mass of exotic quark (T))\footnote{Assuming the (MAC) hypothesis, the most attractive  channel  should satisfy  $\alpha_c(\Lambda_{X})(Y_{L}Y_R) \sim 1$, considering that  $\alpha_c(\Lambda_{X})$  is close to 1, we can roughly estimate that $U(1)_X$ condensation  should occur only for the channel where   $(Y_{L}Y_R) \geq 1$.  Once  $(Y_LY_R = \frac{40}{9})$ the quark T channel is the one leading to the most attractive channel}, then the gauge symmetry $SU(3)_{{}_{L}}\otimes U(1)_{{}_{X}}$ is dynamically broken to $SU(2)_{{}_{L}}\otimes U(1)_{{}_{Y}}$ and the $Z'$, $V^{\pm}$, and $U^{\pm\pm}$ gauge bosons become massive. The $Z'$ mass is given by $M_{Z'} = \frac{g_{X}}{4} F_{\Pi}\left(Y^{T}_L-Y^{T}_R \right)$ , where $F_{\Pi}$ is the pseudoscalar decay constant  and we calculate it using the Pagels-Stokar approximation that is given   by
\be 
 F_{\Pi} = \frac{1}{4\pi^2}\int\!\!\frac{dp^2p^2}{(p^2 + \Sigma^2_{X}(p^2))^2}\!\!\left[\Sigma^2_{X}(p^2) - \frac{p^2}{2}\frac{d\Sigma_{X}(p^2)}{dp^2}\Sigma_{X}(p^2)\right]
\ee
\par Assuming the set of variables described below of Eq.(4), the above equation  together with the definition of $M_{Z'}$, allows to write the   Pagels-Stokar relation as
\be 
1 = \frac{\beta}{\alpha^2}\int^{\left(\!\frac{\Lambda}{M_{Z'}}\!\right)^2\!\!\alpha^2}_{0}\!\!\!\!\!\!\!\!\frac{dxx}{(x + f^2(x))^2}\!\!\left[f^2(x) - \frac{x}{2}\frac{df(x)}{dx}f(x)\right],
\ee 
\noindent where the coefficients $\alpha$ and $\beta$  were defined in the previous section. 
\par According to the description given in \cite{Das} the consistency requirement imposed about the solution  of Eq.(5) is that the mass of the exchanged particle $Z'$(or $\alpha=\frac{M_{Z'}}{M}$)  has to be equal to  $Z'$ mass (or $\alpha$)  obtained using  Eq.(7).
In other words the solutions of the gap equation Eq.(5) are  iteratively improved by starting with a trial guess for the exchanged boson mass and then comparing it with the predicted mass obtained using the Eq.(7). In the Fig I we show   the behavior  of  Eq.(7) assuming the numerical solution of Eq. (5) for $\Lambda=42TeV$, $M_{Z'} =2TeV$ as a function of parameters $\alpha $ and $\beta$. In the table II we  show the results obtained for the dynamically generated masses for heavy  exotic quarks $(T,S,D)$ and exotic techniquarks $(U', D')$ , where we reproduce the results obtained in Ref.\cite{Das} for heavy  exotic quarks . 

\begin{table*}
\caption{The dynamically generated heavy exotic quarks(techniquarks) masses for different $U(1)_{X}$ gauge
coupling $g_{X}$ and cutoff $\Lambda$.}
\label{sphericcase}
\begin{tabular*}{\textwidth}{@{\extracolsep{\fill}}lrrrrrl@{}}
\hline
$\Lambda(TeV)$ & \multicolumn{1}{c}{$M_{Z'}(TeV)$} & \multicolumn{1}{c}{$g_{X}$} & \multicolumn{1}{c}{$\beta$ } & \multicolumn{1}{c}{$M_{T}(TeV)$} & $M_{S,D}(TeV)$  & $M_{U',D'}(TeV)$  \\
\\
\hline
42 & 2.00 & 2.576 & 0.0420 & 5.73 & 0.76  & 0.82   \\ 
42 & 2.50 & 2.576 & 0.0420 & 7.57 & 0.55  & 0.67   \\ 
90 & 3.00 & 2.541 & 0.0409 & 8.22 & 0.92  & 1.12   \\ 
162& 4.00 & 2.526 & 0.0404 & 10.64& 1.22  & 1.41   \\ 
196& 4.50 & 2.523 & 0.0403 & 11.94& 1.34  & 1.47   \\ 
247& 5.00 & 2.514 & 0.0400 & 13.16& 1.52  & 1.75   \\ 
\hline
\end{tabular*}
\end{table*}

\begin{figure}[t]
\begin{center}
\includegraphics[width=0.7\columnwidth]{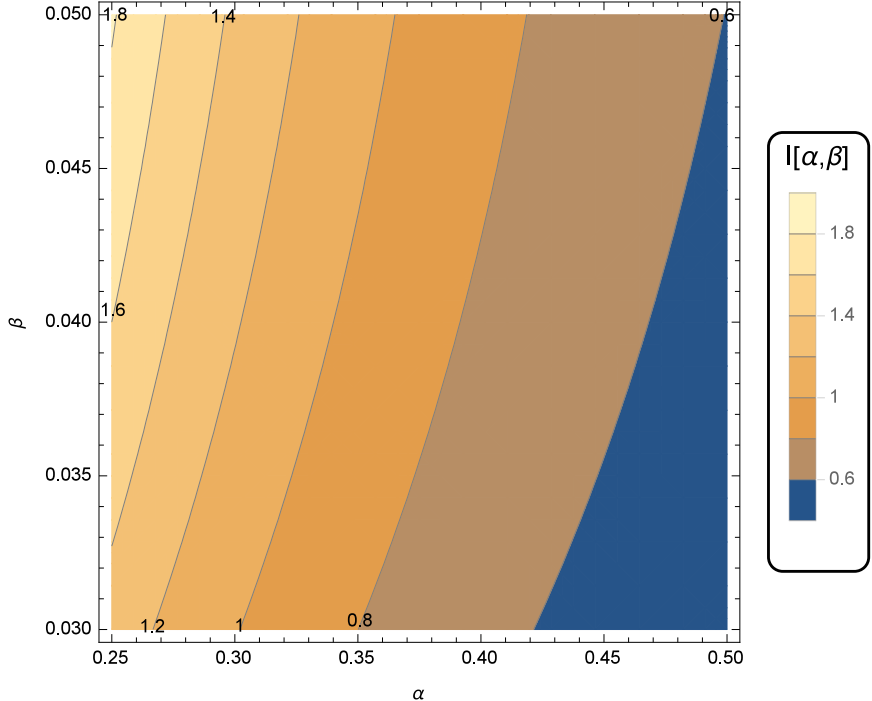}
\caption{Plot of numerical solution of Eqs. (5) and (7) as function of parameters $\alpha $ and $\beta$ to $\Lambda=42TeV$, $M_{Z'} =2TeV$. To  $g_{X} = 2.576$($\beta=0.042$) we get $M_{T}= 5.73TeV$($\alpha=0.349$) which corresponds to the first value obtained for $M_{T}$ described in Ref.\cite{Das}.}
\label{fig1}
\end{center}
\end{figure}

\section{The Extreme Walking Behavior in a 331-TC Model}

\par Theories with large anomalous dimensions ($\gamma_m > 1$) are quite desirable for technicolor phenomenology \cite{takeuchi,tc},  it is  known for a long date that four-fermion interactions are responsible for harder self-energy solutions in non-Abelian gauge theories ($\gamma_m \sim  2$)\cite{takeuchi,miranski}. In the model considered in this work we show that due to the $U(1)_X$  interaction only the exotic techniquarks ($U', D'$)  will acquire   an dynamical mass  $ M_{U'} = M_{D'} = O(TeV)$ at  $\Lambda_{331}\sim O(TeV)$.  The result is that at this energy scale  a bare mass   appears in the (TC) Schwinger-Dyson equation assigned to  exotic techniquarks, ($U', D'$) , what leads to a very "hard" self-energy, or a self-energy of the irregular type, Eq.(1),  only for exotic techniquarks ($U', D'$). In this section considering a four-fermion approximation  for $U (1)_X$ interaction associated to these techniquarks we will show that the results for $M_{U'} = M_{D'} = O(TeV)$ are of the same order as obtained in the previous section, in this case $F_{\Pi}$ equation is given by
\br 
F^2_{\Pi} = \frac{1}{4\pi^2}\int\!\!\frac{dp^2p^2 M^2_{X}}{(p^2 + M^2_{X})^2} 
\er 
\noindent where $M_{X} =  M_{U'} = M_{D'}$ or $M_{T}$. As in the previous section, the equation above together with the definition of $M_{Z'}$, allows us to write 
\br 
1 \approx \frac{\beta}{\alpha^2}\int\!\!\frac{dp^2p^2}{(p^2 + M^2_{X})^2}. 
\er
\par For a similar choice of parameters, used in the previous section, for example  $\Lambda=42$ TeV, $\beta = 0.042$ and $M_{Z'}=2.5$ TeV we obtain $M_T \approx 6.3 $TeV  and $M_{U'} = M_{D'}\approx 0.6$TeV so that the contribution due  $U(1)_X$   to the mass of exotic techniquarks can be approximated by a four-fermion  interaction and the exotic techniquarks exhibit a self-energy behavior of the irregular type. 
\par To illustrated  the  extreme walking behavior exhibited  only by the exotic technifermions  we consider the full gap equation for the "exotic techniquark U'" that contains the sum of two contributions, the $U(1)_X$ interaction and TC interaction,  and  we consider the presence of dynamically massive technigluons. The problems for chiral symmetry breaking in this case have been discussed recently, where confinement may
play an important role\cite{agn,rdn,fdn}. In this work we consider that  technigluons acquire a dynamical mass along the line of QCD as  proposed by  Cornwall\cite{Cornwall} many years  ago and the dynamical technigluon mass  behaves as\cite{agn,rdn}
\be 
m^2_{tg}(k^2) = \frac{m^4_{tg}(0)}{k^2 + m^2_{tg}(0)},  
\ee 
\noindent   with  $m_{tg}(0) \approx \Lambda_{TC}$ and the technifermion dynamical mass will be given by
\br
\!\!\!\!\! M_{TC}(p^2) = \frac{C_2}{(2\pi)^4}\int \!\!\frac{K(g,p)d^4k}{[k^2+M^2_{TC}(k^2)]},
\label{eqn1}
\er
\noindent where 
$$
K(g,p)=\frac{3\bar{g}^2_{tc}(p-k)M_{TC}(k^2)}{(p-k)^2+m_{tg}^2((p-k)^2)}
$$ 
\noindent $M_{TC}(p^2)$ is the dynamical techniquark mass,  and $C_2$ is the Casimir operator for the  technifermionic representation with effective TC coupling ${\bar{g}}_{tc}$,   given by
\be
{\bar{g}}^2_{tg}(p^2)= \frac{1}{b_{TC} \ln[(p^2+4m^2_{tg}(p^2))/\Lambda_{TC}^2]} \, ,
\ee
\noindent  in this expression $b_{TC}$ is the first $\beta_{TC}$ function coefficient and we consider the Landau gauge. Similarly to the previous section the  integral equation  described by  Eq.(11) can be transformed into a differential equation for $h(x)$  introducing  the  new variables $x = \frac{p^2}{M^2_{TC}(0)}$,   with $h(x) = \frac{M_{TC}(x)}{M_{TC}(0)}$ and  $\delta=\frac{m_{tg}(0)}{M_{TC}(0)}$ we obtain  
\be 
h''(x) + \frac{2}{x + \delta^2}h'(x) + \frac{3C_2g^2_{TC}(x)}{16\pi^2(x +\delta^2)^2}\frac{xh(x)}{(x + h^2(x))} = 0. 
\ee 

\par The full gap equation for the "exotic techniquark U'" can be written as 
\be 
M_{U'}(x)  =  M_{TC}(0)h(x) + M_{X}(4f), 
\ee 
\noindent in the Fig.(2a)[Blue curve] we show the behavior of $M_{TC}(x) = \frac{\langle \bar{U}{U}\rangle}{\Lambda^2}$,  that corresponds to the  dynamical mass generated for the the usual techniquarks $(U,D)$ obtained with Eq.(13). In the Fig.(2b)[Red curve] we include   the  contribution of $U(1)_X$ effective four-fermion interaction and we show the behavior of the dynamical mass generated for the exotic techniquarks $(U',D')$, which is described by $M_{U'}(x) = \frac{\langle \bar{U'}{U'}\rangle}{\Lambda^2}$. 
\begin{figure}[ht]
\begin{center}
\includegraphics[width=0.9\columnwidth]{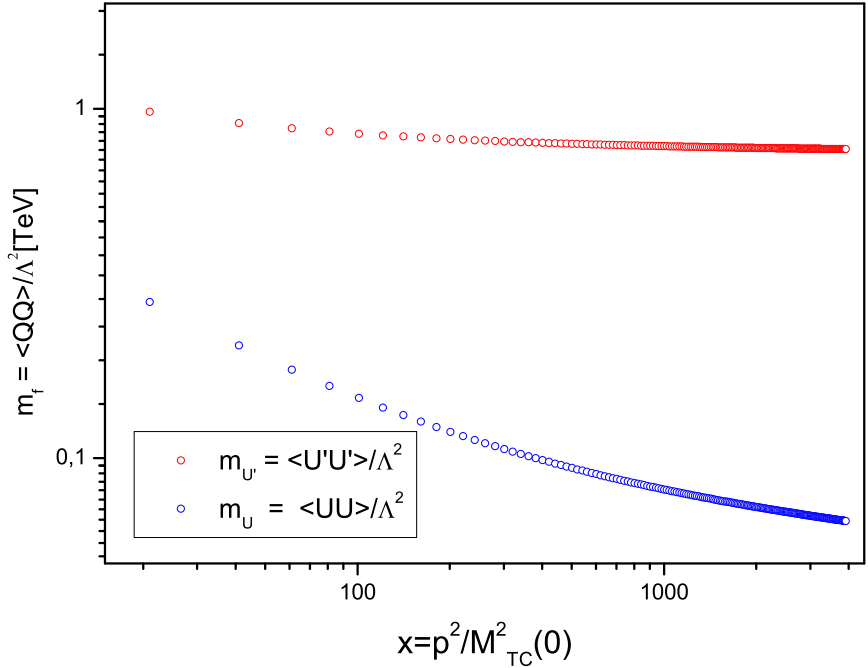}
\caption{In this figure  we plot the behavior  of $M_{TC}(x) = \frac{\langle \bar{U}{U}\rangle}{\Lambda^2}$ obtained with Eq.(13) to  $N_{TC} = SU(2)_{TC}$ and $n_{TF} = 6$ fig.(2a)[curve in Blue] and in the fig.(2b)[curve in Red] we show the behavior of dynamical mass generated for the exotic techniquarks $(U',D')$, given by $M_{U'}(x) = \frac{\langle \bar{U'}{U'}\rangle}{\Lambda^2}$.  }
\label{fig1}
\end{center}
\end{figure}

\newpage
\section{Conclusions}
\par   In the Refs.\cite{us1,us2,us3,twoscale} we discussed the possibility of  obtaining a  light composite TC scalar boson,  this result is a direct consequence  of extreme walking (or quasi-conformal) technicolor theories, where the  asymptotic self-energy  behavior  is described by Eq.(1).  The extreme walking technicolor can be obtained in  three different ways and in this paper we propose an scheme to obtain the quasi-conformal behavior based on an extension of the electroweak sector of the standard model,  in the so called 331-TC model ($SU(N)_{TC}\otimes SU(3)_{{}_{L}}\otimes SU(3)_{{}_{c}}\otimes U(1)_{{}_{X}}$), where  the  exotic quark $T$  introduced  will form a  $U(1)_X$ condensate $(\langle \bar{T}T \rangle)$ breaking  $SU(3)_{{}_{L}}\otimes U(1)_X$ to  electroweak symmetry that is broken  by an usual  technifermion condensate. As the comment  presented in the introductory section, 3-3-1 models predicted new massive weak gauge bosons, $Z'$  and  $V^{\pm}$, and this model has the necessary requirements to explain the reported diboson anomaly. 
\par Following  the same procedure described in Ref.\cite{Das}, the solutions of the gap equation, Eq.(5), were  iteratively improved by starting with a trial guess for the exchanged boson mass and then comparing it with the predicted mass obtained using the Eq.(7). In the table II we  show the results obtained for the dynamically generated masses of heavy  exotic quarks $(T,S,D)$ and exotic techniquarks $(U', D')$ , where we reproduce the results obtained in  Ref.\cite{Das} for heavy  exotic quarks.  In Ref.\cite{an3} we discuss a mechanism for
 the dynamical mass generation, including the mass generation for the t quark, in the case of grand unified theories that incorporate  quarks and techniquarks. We expect that a similar mechanism to the one described in \cite{an3} can be developed and incorporated in the present model.
 In the section 4 we show that the results obtained in the table II, ($M_{U'} = M_{D'} = O (TeV) $),  are of the same order as the ones obtained with a four-fermion approximation for the $U (1)_X$ interaction and in the Fig.2 [Red Curve ] we show the extreme walking behavior displayed only  by exotic techniquarks ($U', D'$) due to their strong $U(1)_X$ interaction because $Y^{U'}_{L}Y^{U'}_{R}=3$ (see  table I). 
 The exotic technifermions  have two different  energy scales and the 331-TC model corresponds to an example of  two-scale Technicolor (TC) model, in the Ref.\cite{twoscale}  we considered the possibility of a light TC composite boson arising from mass mixing between a relatively light and  heavy composite scalar singlets from a see-saw mechanism  expected to occur in two-scale TC model. We  emphasize that the see-saw mechanism expected to occur in this model is not exactly the same one described in the Ref[16], in the model proposed the extreme walking behavior displayed  by exotic techniquarks is due to their strong $U(1)_X$ interaction and in this case in  Eq.(1), $\mu\approx\Lambda_2 =\Lambda_{331}$, and $\Lambda_2$ it is  not generated by the  TC sector.  In order to provide a example we consider  the $SU(2)_{TC}$ case with the  technifermionic content showed in[6], where $n_{TF} = 6$. With this we obtain $m_{\phi_1} \sim 124$ GeV,  $m_{\phi_2} \sim 8.3 $TeV to $\Lambda_{331}=3TeV$ and $\Lambda_{ETC}\approx 90$TeV.  Therefore, in this scenario it is possible to obtain a  light composite TC scalar boson and include the necessary requirements to explain the diboson anomaly,  the determination of the scalar spectrum  for composite Higgs bosons of this model will be presented in detail  in a future work.



\begin{acknowledgments}
I thank A. A. Natale for useful discussions. This research was partially supported by the Conselho Nacional de Desenvolvimento Cient\'{\i}fico e Tecnol\'ogico (CNPq)  by grant 442009/2014-3.  
\end{acknowledgments}


\end{document}